\documentclass[aps,pra,twocolumn,amsmath,amssymb,superscriptaddress,reprint]{revtex4-1}

\usepackage[utf8]{inputenc}
\usepackage{graphicx}
\usepackage[colorlinks=true,citecolor=blue]{hyperref}
\usepackage{units}
\usepackage{bm}

\newcommand{\bra}[1]{\langle #1|}
\newcommand{\ket}[1]{|#1\rangle}
\DeclareMathOperator{\Tr}{Tr}

\renewcommand{\vec}[1]{\mathbf{#1}}

\newcommand{\kBT}{k_\text{B}T}
\newcommand{\tL}{\text{L}}
\newcommand{\tR}{\text{R}}

\begin{document}
\title{Relaxation dynamics in double-spin systems}
\author{Philipp Stegmann}\email{philipp.stegmann@uni-due.de}
\affiliation{Theoretische Physik, Universit\"at Duisburg-Essen and CENIDE, D-47048 Duisburg, Germany}
\author{Jürgen König}
\affiliation{Theoretische Physik, Universit\"at Duisburg-Essen and CENIDE, D-47048 Duisburg, Germany}
\author{Björn Sothmann}
\affiliation{Theoretische Physik, Universit\"at Duisburg-Essen and CENIDE, D-47048 Duisburg, Germany}
\date{\today}

\begin{abstract}
We consider the relaxation dynamics of two spins coupled to a common bosonic bath. The time evolution is simulated by a generalized master equation derived within a real-time diagrammatic approach. Interference effects due to the coherent coupling to the common bath give rise to characteristic features in the relaxation dynamics after a quench or during a periodic external driving. 
In particular, we find that the long-time behavior during periodic driving depends sensitively on the initial state as well as on system parameters such as coupling asymmetries.
When coupled to more than a single reservoir, the interference effects can lead to a cooling mechanism for one of the bosonic reservoirs.
\end{abstract}

\maketitle
\section{\label{sec:intro}Introduction}
With the ongoing miniaturization of electronic circuits it has become increasingly more important to fabricate, control, and manipulate systems at the nanoscale.
Recent years have shown tremendous progress in this regard that has led, e.g., to exciting developments such as arrays of superconducting qubits~\cite{clarke_superconducting_2008,martinis_superconducting_2009,devoret_superconducting_2013} and microwave cavities~\cite{schoelkopf_wiring_2008,wendin_quantum_2017,gu_microwave_2017} for quantum computing.
Nanoscale systems based on spins are of particular interest as their quantum-mechanical nature is inherently important. They can be realized in a number of different ways such as NV centers in diamond~\cite{doherty_nitrogen-vacancy_2013,schirhagl_nitrogen-vacancy_2014}, magnetic nanoparticles~\cite{wernsdorfer_classical_2001}, single molecular magnets~\cite{sessoli_magnetic_1993,gatteschi_molecular_2006,bogani_molecular_2008} and individual magnetic atoms~\cite{heinrich_single-atom_2004,hirjibehedin_spin_2006,hirjibehedin_large_2007,meier_revealing_2008,ternes_spin_2015}, nuclear spins in semiconductors~\cite{chekhovich_nuclear_2013,scheller_possible_2014}, and electron spins in semiconductor quantum dots~\cite{petta_coherent_2005,hanson_spins_2007,warburton_single_2013}. The various spin systems offer promising prospects for applications in high-density memories~\cite{kalff_kilobyte_2016,natterer_reading_2017} and for spin-based quantum computation~\cite{loss_quantum_1998}.

In general, quantum systems are not totally isolated but are coupled to some environmental degrees of freedom. For spin systems, the coupling to the environment gives rise to energy relaxation and dephasing which have both been studied extensively in the framework of the spin-boson model~\cite{leggett_dynamics_1987,leggett_erratum:_1995,prokofev_theory_2000} where a central spin is coupled to an ensemble of harmonic oscillators that act as bath of noninteracting bosons.
While energy relaxation and dephasing are detrimental for quantum computing and in memory applications, the coupling of a spin to different baths that are not in thermodynamic equilibrium with each other gives rise to interesting new effects such as the possibility of driving the spin into nonequilibrium states, of heat transport across the spin~\cite{ruokola_thermal_2011,nicolin_non-equilibrium_2011,chen_dynamic_2013,wang_nonequilibrium_2015,wang_unifying_2017,yamamoto_heat_2018}, and even rectification of such heat currents in nonlinear response~\cite{segal_spin-boson_2005}.

\begin{figure}
    \includegraphics[width=\columnwidth]{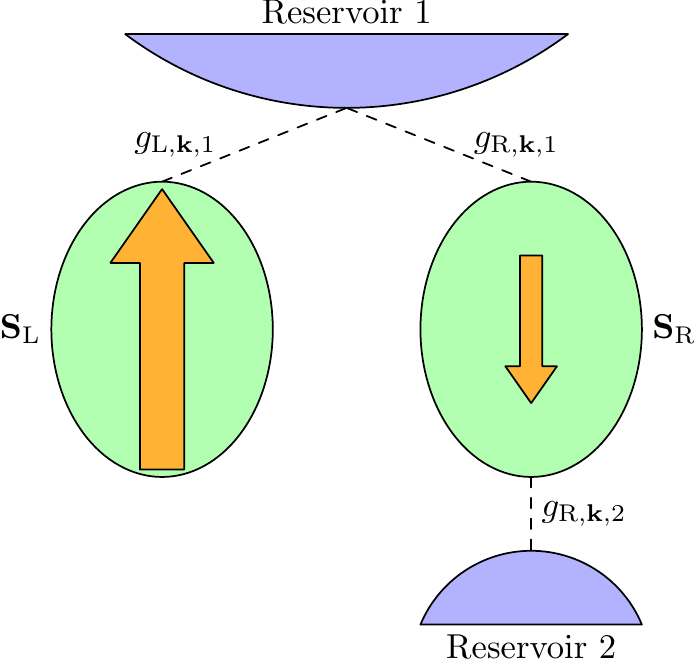}
	\caption{\label{fig:model}Sketch of the system. Two spins that are each characterized by their spin $\vec S_i$, $i=\text{L,R}$, are coupled to a common bosonic reservoir with coupling constant $g_{i,\vec k,1}$. In addition, the second spin couples to an additional bosonic reservoir with coupling constant $g_{\text{R},\vec k,2}$.}
\end{figure}

So far, most studies of the spin-boson model have focused on setups where a single central spin is coupled to one or several bosonic baths. Here, we go beyond this paradigm by considering an extended quantum system consisting of two spins coupled to bosonic reservoirs, see Fig.~\ref{fig:model}. As a new ingredient, the nonlocal nature of the spin system allows for the coherent coupling of the different spins to one common reservoir. As was pointed out recently in Refs.~\cite{hama_relaxation_2018,hama_negative-temperature-state_2018}, this coherent coupling can give rise to new and interesting effects such as the possibility of relaxing certain initial states not into the ground state but rather into an excited state. When considering only the smaller of the two spins in this nonequilibrium state, one observes an occupation inversion that can, in principle, be interpreted in the framework of negative temperatures~\cite{purcell_nuclear_1951,ramsey_thermodynamics_1956,dunkel_consistent_2014}. Physically, this counterintuitive relaxation dynamics is based on the conservation of angular momentum. If both spins couple equally to the common reservoir, the total angular momentum of the spin system is conserved during relaxation. Hence, if the initial state has a different total angular momentum than the ground state, relaxation to the lowest energy state is not possible and the system instead ends up in the lowest energy state compatible with angular momentum conservation.

Here, we extend the analysis of Hama \textit{et al.}~\cite{hama_relaxation_2018,hama_negative-temperature-state_2018} in several regards. First of all, on a technical level we make use of a real-time diagrammatic approach~\cite{konig_zero-bias_1996,konig_resonant_1996,schoeller_transport_1997,konig_quantum_1999,timm_tunneling_2008,koller_density-operator_2010} to derive a generalized master equation that determines the relaxation dynamics of the spin system. The real-time diagrammatics is tailored to deal with arbitrary interactions within the quantum system. Hence, we do not have to resort to any mean-field approximation in evaluating spin expectation values. Furthermore, our approach allows for a systematic expansion in the system-reservoir coupling which accounts for level renormalization effects that impact the system dynamics even to lowest order in the tunnel couplings~\cite{konig_interaction-driven_2003,braun_theory_2004,sothmann_influence_2010,kamp_phase-dependent_2019}.
Second, we do not only consider the dynamics of the system after a quench of its parameters but we also consider the response to a periodic external driving. We find that the long-time behavior reached under periodic driving can be very different from the final state that is reached after a simple quench under otherwise identical conditions.
Finally, we explore a larger range of parameter space by allowing for asymmetric couplings of the two spins to the common reservoir. This breaks total angular momentum conservation and, thus, the mechanism that gives rise to the population inversion. Nevertheless, we find that the system dynamics still remains nontrivial in this case. Furthermore, we also introduce the coupling of one spin to an additional second reservoir which also gives rise to a breaking of angular momentum conservation.

The paper is organized as follows. We present the system and its theoretical description in Sec.~\ref{sec:model}. The results for driving the setup with a periodically modulated external magnetic field are discussed in Sec.~\ref{ssec:periodic} while the dynamics following a quench is analyzed in Sec.~\ref{ssec:quench}. Finally, we present conclusions in Sec.~\ref{sec:con}.

\section{\label{sec:model}Model}
We consider a setup consisting of two spins with spin $\vec S_i$, $i=\tL,\tR$, coupled to bosonic reservoirs. In particular, both spins couple coherently to one common reservoir while the second spin may additionally couple to a separate bath, see Fig.~\ref{fig:model}. While in principle both spins can take arbitrary values, here we will focus on the case where $S_\tL\gg S_\tR$. In the following, we will consider the limiting case $S_\tR=1/2$. Nevertheless, our results remain qualitatively valid for larger values of $S_\tR$.
The two spins are subject to a time-dependent external magnetic field $B(t)$ along the $z$ direction which we assume to be identical for both spins. The spins are thus described by the Hamiltonian
\begin{equation}
	H_\text{spin}=B(t) \sum_i S^z_i.
\end{equation}
The reservoirs $j=1,2$ are modeled in terms of noninteracting bosons with dispersion relation $E_{\vec k,j}$ as
\begin{equation}
	H_{\text{res},j}=\sum_{\vec k} E_{\vec k,j}b_{\vec k,j}^\dagger b_{\vec k,j}.
\end{equation}
We assume that both reservoirs have the same temperature $T$.
Finally, the coupling between the spins and the reservoirs is given by the Hamiltonian
\begin{equation}\label{eq:Hint}
	H_\text{int}=\sum_{i,j,\vec k}\left( g_{i,\vec k,j} S^{\sigma}_i b_{\vec k,j}+ g^*_{i,\vec k,j} S^{-\sigma}_i b_{\vec k,j}^\dagger \right)\, ,
\end{equation}
with $\sigma=\text{sgn}[B(t)]$ characterizing the direction of the magnetic field.
It describes processes where a boson is created (annihilated) in bath $j$ while the $z$ component of spin $i$ is decreased (increased) by $\hbar$ in the energetically favored direction. The coupling constants $g_{i,\vec k, j}$ are related to the energy-dependent spectral functions $J^{(j)}_{ii'}(\omega)=\frac{2\pi}{\hbar}\sum_{\vec k} g_{i,\vec k,j} g_{i',\vec k,j}^* \delta(\omega-E_{\vec k,j})=\gamma_{ii'}^{(j)}\omega^\alpha/(B\omega_\text{c}^{\alpha-1}) e^{-\omega/\omega_\text{c}}$, where $\omega_\text{c}$ denotes a cutoff energy and $B$ is the time-averaged absolute value of the magnetic field. The exponent $\alpha$ is connected to the energy dependence of the density of states $\sum_{\vec k}\delta(\omega-E_{\vec k,j})$. In the following, we will assume an Ohmic bath with $\alpha=1$. The first reservoir is coupled to both spins which yields separate tunnel coupling strengths ($\gamma^{(1)}_{\tL\tL}$,$\gamma^{(1)}_{\tR\tR}$) and also interference contributions that scale with $\gamma^{(1)}_{\tL \tR}=\gamma^{(1)}_{\tR \tL}= \sqrt{\gamma^{(1)}_{\tL\tL} \gamma^{(1)}_{\tR\tR}}$. The latter couple the dynamics of both spins coherently. If neglected, the spins evolve independently of each other as if each spin is coupled to its own reservoir. The second reservoir couples only to the right spin and, therefore, only $\gamma^{(2)}_{\tR\tR}$ must be taken into account. In this paper, we choose $1/(\gamma^{(1)}_\text{LL} + \gamma^{(1)}_\text{RR})$ and $k_\text{B}T$ as units of time and energy, respectively. 

In order to describe the time evolution of the system, we extend a real-time diagrammatic technique that has originally been developed to describe electronic transport in strongly interacting quantum dot systems~\cite{konig_zero-bias_1996,konig_resonant_1996,schoeller_transport_1997,konig_quantum_1999,timm_tunneling_2008,koller_density-operator_2010}. The key idea behind this approach is to integrate out the noninteracting bath degrees of freedom and to describe the remaining quantum system in terms of its reduced density matrix $\rho$ with matrix elements $\rho_{\chi_1,\chi_2}=\langle \chi_1|\rho|\chi_2\rangle$ where $\chi_{1,2}$ are eigenstates of $H_\text{spin}$. For the system at hands, we can either choose the basis spanned by the total spin of the system $S_\text{tot}$ and its $z$ component $S_\text{tot}^z$ or the basis spanned by the $z$ component of both spins, $S^z_\text{L}$ and $S^z_\text{R}$. While both descriptions are completely equivalent to each other, some aspects of the spin dynamics can be understood more easily in one of the two bases.
The real-time diagrammatics allows for an exact treatment of any type of interactions in the quantum system in combination with a systematic expansion in the tunnel couplings to the reservoirs. 
The time evolution of the reduced density matrix is governed by a generalized master equation of the form
\begin{multline}\label{eq:master}
	\frac{d}{dt}\rho_{\chi_1,\chi_2}(t)=-i(E_{\chi_1}-E_{\chi_2})\rho_{\chi_1,\chi_2}(t)\\+\sum_{\chi'_1,\chi'_2}\int_0^t dt' W_{\chi_1,\chi_2;\chi'_1,\chi'_2}(t-t')\rho_{\chi'_1,\chi'_2}(t').
\end{multline}
The first term on the right-hand side describes the coherent evolution of the system in the absence of any coupling to the bath. The second term arises from the dissipative system-bath coupling. The generalized transition rates $W_{\chi_1,\chi_2;\chi'_1,\chi'_2}$ can be evaluated as irreducible blocks of the quantum system's propagator on the Keldysh contour. In this work, we evaluate the transition rates to first order in the system-reservoir coupling strengths $\gamma^{(j)}_{ii'}$. Once we have obtained the time-dependent reduced density matrix, we can evaluate spin expectation values in a standard way as $\langle S^z_i\rangle=\Tr S^z_i \rho$. The heat currents flowing into the reservoirs are given by the expectation value of the heat current kernel $W_Q$ which is obtained from the generalized transition rates $W$ by multiplying each rate with the heat transferred from the spins into the reservoir during the associated transition.

The generalized master equation accounts for the dynamics of both, diagonal as well as off-diagonal elements of the reduced density matrix. Here, we consider initial density matrices $\rho(t=0)$ that are diagonal in $S_\text{tot}^z$. The coupling Hamiltonian $(\ref{eq:Hint})$ does not generate coherent superposition between states with different $S_\text{tot}^z$. In contrast, coherent superposition between states with different $S_\text{tot}$ can arise even if they are absent at $t=0$. Since the eigenenergies $E_\chi=\bra{\chi}H_\text{spin}\ket{\chi}$ do not depend on the total spin, the first term on the right-hand side of Eq.~(\ref{eq:master}) drops out. The coherent dynamics is completely governed by the dissipative coupling to the reservoirs. The solution of the master equation is presented in detail in Appendix \ref{ap:sol}.

Importantly, the generalized transition rates that involve coherences in general have a real as well as an imaginary part. While the former describes the rate of transition, the latter determines the frequency of coherent oscillations between states with different $S_\text{tot}$. It arises from virtual transitions and can be expressed (up to a Clebsch-Gordan coefficient) by the principal value integral
\begin{equation}
	B^{(j)\pm}_{\text{ex},i i'}=\mathcal{P}\int_{0}^\infty d\omega\frac{J^{(j)}_{ii'}(\omega)n^{\pm}(\omega)}{B-\omega},
\end{equation}
where $n^{+}(\omega)=[\exp(\omega/\kBT)-1]^{-1}$ denotes the Bose function and $n^-(\omega)=n^{+}(\omega)+1$. Assuming that the two total spin values form a pseudospin-1/2, the oscillations can be seen as an electron spin precessing around an effective exchange field~\cite{konig_interaction-driven_2003,braun_theory_2004,sothmann_influence_2010,misiorny_underscreened_2012,misiorny_spintronic_2013,hell_spin_2015,weiss_spin_2015,crisan_harnessing_2016,stegmann_coherent_2018}. Such an analogy has also been observed in the context of quantum transport though spatially localized~\cite{kamp_phase-dependent_2019,donarini_coherent_2019,ho_counting_2019} and nonlocalized orbitals~\cite{wunsch_probing_2005}. However, in contrast to the electronic case, here, the exchange field grows with the cutoff energy, i.e., $B^{(j)+}_{\text{ex},i i'}+B^{(j)-}_{\text{ex},i i'} \propto \omega_{\text{c}}$. We assume that the cutoff energy is the largest energy scale of the system. Thus, the frequency of coherent oscillations is much larger than the real part of the transition rates. The specific values of $B$ and $k_\text{B}T$ are not relevant for the coherent oscillations.

We emphasize that a proper treatment of the exchange field is needed to obtain a correct description of the time evolution of the system. This becomes obvious when analyzing the time evolution in the two spin bases mentioned above. Identical results are obtained only if the exchange field is included in the calculation.

\section{\label{sec:res}Results}
One of the most interesting properties of the double-spin system is the presence of \textit{multistability}~\cite{schaller_counting_2010, buvca_note_2012, manzano_symmetry_2014, albert_symmetries_2014, thingna_dynamical_2016,stegmann_inverse_2017,caspel_symmetry_2018,thingna_magnetic_2019} in the limit of low temperatures. If the magnetic field becomes time-independent, the system relaxes into one or a mixture of different states with $\dot \rho(t)=0$~\footnote{These states can be obtained as eigenvectors fulfilling $\bm{W \rho}=0$ and $\text{Tr}[\bm{\rho}]=1$. The matrix $\bm{W}$ is defined in the Appendix~\ref{ap:sol}}. The direction of the magnetic field determines the $z$ components $S^z_{\text{L/R}}$ of each possible state. For $B(t) \lessgtr 0$, the ground state $\ket{S^z_\tL=\pm S_\tL$, $S^z_\tR= \pm S_\tR}$ is stable. The form of the additional states is determined by the system-reservoir coupling strengths and the size of the spins $S_{\text{L/R}}$. The mixture of states that the system relaxes into depends on the initial state. In the following, we elucidate this behavior in more detail. We analyze the spin dynamics as well as the heat flowing out of the reservoirs. In particular, we are going to discuss two scenarios namely (i) a periodic driving of the system and (ii) its quench dynamics.

\subsection{\label{ssec:periodic}Periodic driving}
We consider the system in the special case where the spins are coupled to the common reservoir only and the coupling to the second reservoir vanishes, $\gamma^{(2)}_{\tR \tR}(\omega)=0$. The system is prepared in either of two initial states. In the first state, both spin are aligned parallel and point along the positive $z$ axis, $\ket{\psi_\text{P}}=\ket{S^z_\tL=S_\tL$, $S^z_\tR=S_\tR}$. In the second state, the two spins are aligned antiparallely such that $\ket{\psi_\text{AP}}=\ket{S^z_\tL=S_\tL$, $S^z_\tR=-S_\tR}$. Subsequently, the setup is subject to a periodic switching of the magnetic field direction with period $\mathcal{T}$ such that $B(t)=+B$ during the first half period and $B(t)=-B$ during the second half period. The length of each half period is sufficient for the system to relax into a state which is stable until the magnetic field switches.

\paragraph{Symmetric coupling}
\begin{figure}
    \includegraphics[width=\columnwidth]{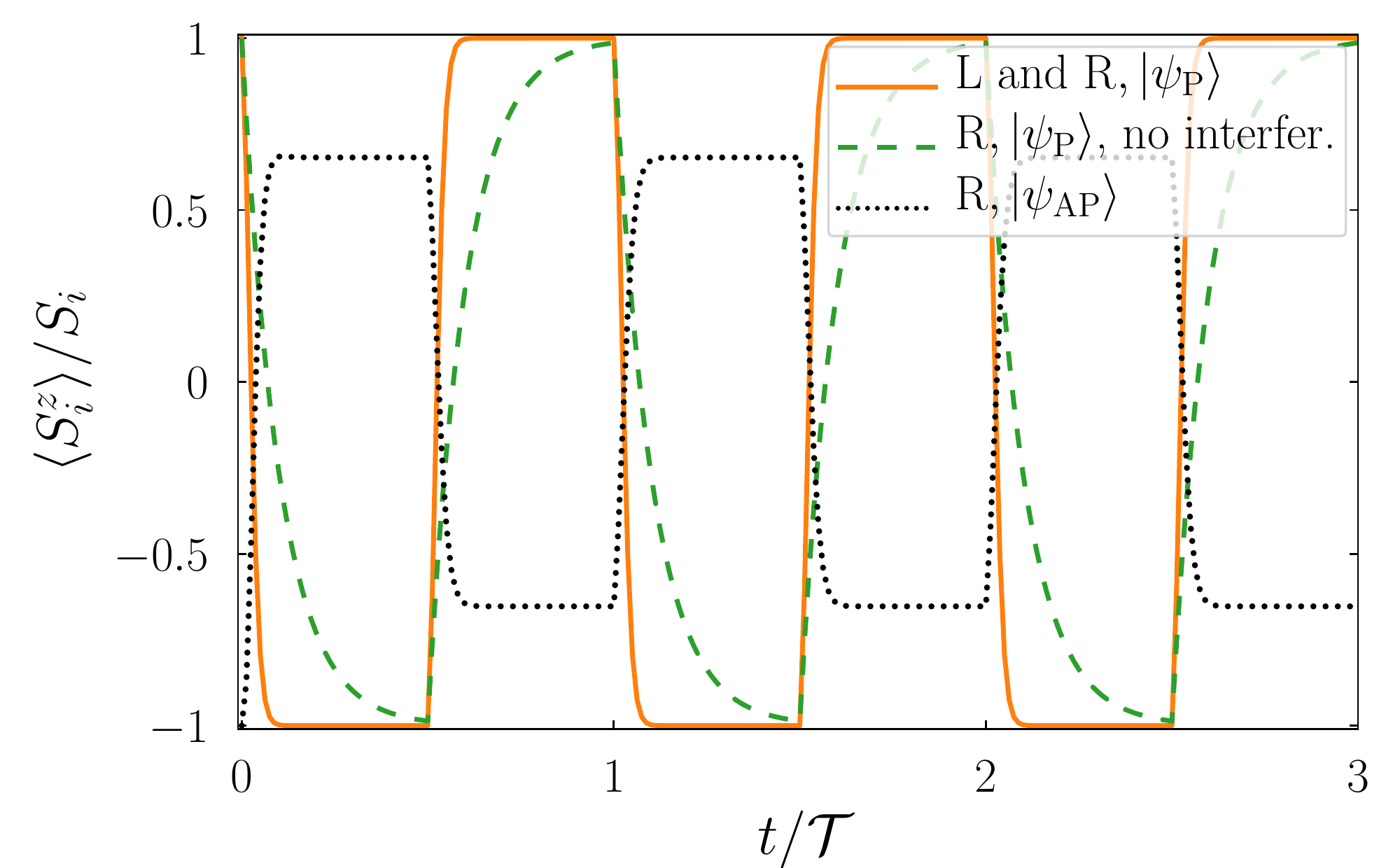}
	\caption{\label{fig:symmetric}Time evolution of $\langle S^z_\tL\rangle $ and $\langle S^z_\tR\rangle $ for symmetric couplings $\gamma^{(1)}_{\tL \tL}=\gamma^{(1)}_{\tR \tR}$. The total spins are $S_\tL=5$ and $S_\tR=1/2$. The magnetic field of strength $B=10\, k_\text{B} T$ changes its sign twice each period $\mathcal{T}=10/ (\gamma^{(1)}_{\tL \tL}+\gamma^{(1)}_{\tR \tR})$. The cutoff energy is $\omega_\text{c}=10^2 k_\text{B} T$.}
\end{figure}

To start with, we consider the case where both spins couple symmetrically to the common reservoir, i.e., $\gamma^{(1)}_{\tL \tL}=\gamma^{(1)}_{\tR \tR}$.
If the system is prepared in the parallel initial state $\ket{\psi_\text{P}}$, the dynamics of the spin system is rather trivial, cf. Fig.~\ref{fig:symmetric}. Whenever the magnetic field switches, both spins (orange curve) flip such that the system relaxes to the new ground state, provided $\kBT\ll B$. The dynamics of this process can be understood most easily in the basis of the total spin of the two spins and its $z$ component. In order to switch from the state $S^z_\text{tot}=+S_\text{tot}$ to the state $S^z_\text{tot}=-S_\text{tot}$, the system has to create $2S_\text{tot}+1$ bosons. The rate for this creation process scales as $|\langle S_\text{tot},S^z_\text{tot}-1|S^-_\text{tot}|S_\text{tot},S^z_\text{tot}\rangle|^2\propto S_\text{tot}^2$. Hence, the switching time scales as $S_\text{tot}^{-1}$, i.e., the spins switch faster the larger they are. Due to the symmetric couplings this switching affects the spin of the left and right spin equally. This effect relies crucially on the interference-related couplings. If we neglect the coherent coupling, $\gamma^{(1)}_{\tL \tR}$=$\gamma^{(1)}_{\tR \tL}$=0, the spins evolve independently of each other. In this case, the switching time of $\langle S^z_\tR \rangle$ is drastically reduced, cf. the dashed green curve in Fig.~\ref{fig:symmetric}.

The dynamics of the system becomes more interesting if it is initially prepared in the antiparallel configuration $\ket{\psi_\text{AP}}$ (dotted black curve in Fig.~\ref{fig:symmetric}).
Right after the initial switching of the magnetic field at $t=0$, the left spin is in its highest excited state while the right spin is in its ground state. Naively, one might expect the left spin to relax to its ground state while the right spin stays in its ground state. Interestingly, this does not happen. While the large spin indeed relaxes to the ground state, the small spin flips from the ground state to some excited state such that it exhibits a population inversion with the excited state having a higher occupation probability than the ground state. Therefore, if only the right spin is considered, it seems as if the system has reached a negative-temperature state. We remark, however, that the state of either spins is not a thermal but rather a generic nonequilibrium state. This counterintuitive behavior has been noticed previously in Refs.~\cite{hama_relaxation_2018,hama_negative-temperature-state_2018}. It arises because the Hamiltonian~(\ref{eq:Hint}) takes the form~$H_\text{int}=\sum_{\vec k}\left( g_{\tL,\vec k,1} {S^\sigma_\text{tot}} b_{\vec k,1}+ g^*_{\tL,\vec k,1} S^{-\sigma}_\text{tot} b_{\vec k,1}^\dagger \right)$ for symmetric couplings ($g_{\tL,\vec k,1}=g_{\tR,\vec k,1}$) such that the total angular momentum is conserved. Hence, a system prepared in the initial state $\ket{\psi_\text{AP}}$ with $S_\text{tot}\neq(S_\tL+S_\tR)$ cannot relax to the ground state with $S_\text{tot}=S_\tL+S_\tR$ but rather ends up in an excited state which is stable until the magnetic field switches again. In the limit $S_\tR=1/2$ and large $S_\tL$, this excited state takes the form $\ket{S^z_\tL=\pm S_\tL$, $S^z_\tR= \mp 1/2}$ for $B(t) \lessgtr 0$, i.e., the probability to find the right spin in its ground state vanishes.

We remark that in order to describe the nontrivial dynamics of the second spin correctly, it is again crucial to properly account for the interference-related couplings. If neglected, $\langle S^z_\tR \rangle$ follows the direction of the magnetic field as indicated by the dashed green curve in Fig.~\ref{fig:symmetric}.

\paragraph{Asymmetric coupling}
\begin{figure}
    \includegraphics[width=\columnwidth]{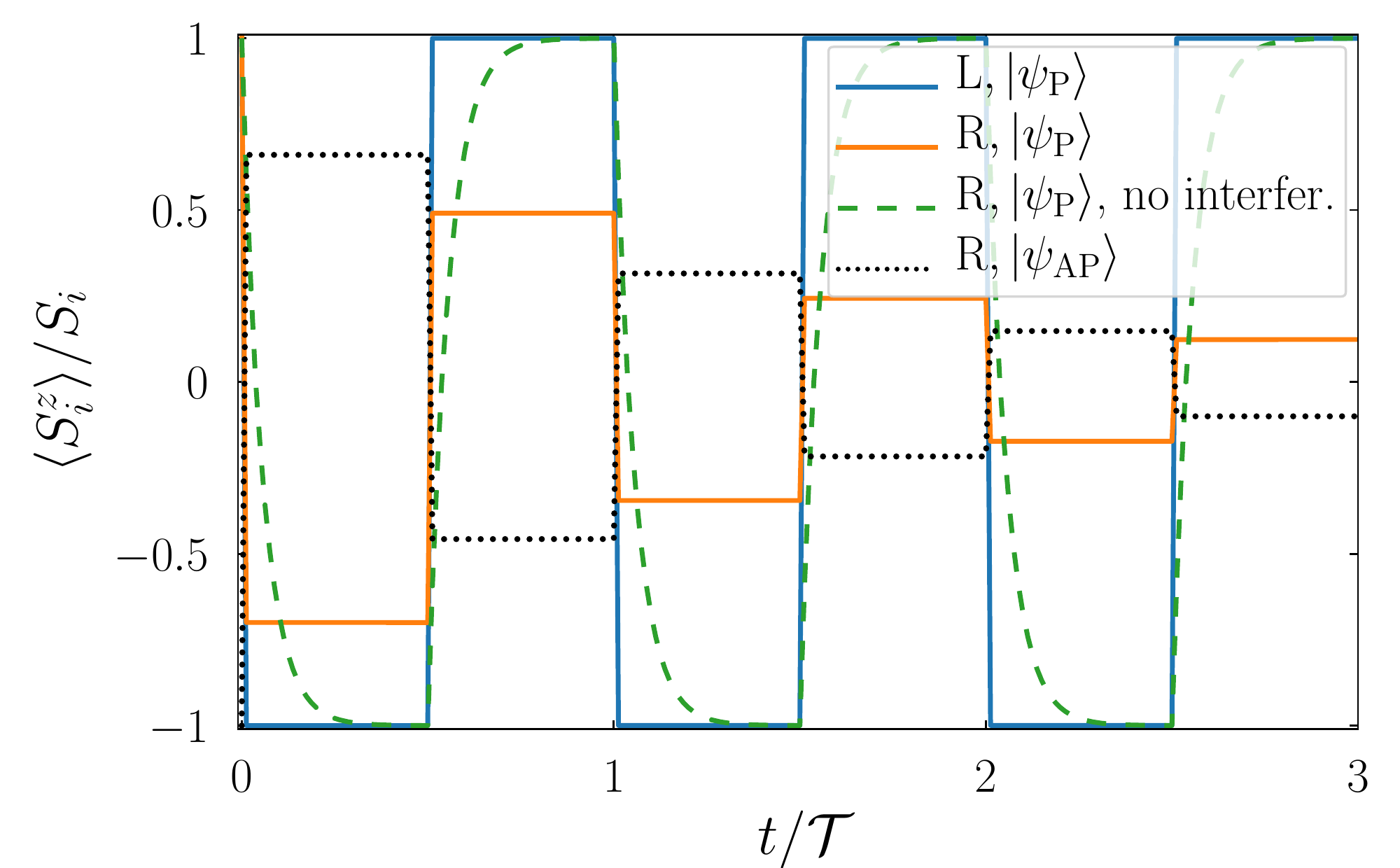}
	\caption{\label{fig:asymmetric1}Time evolution of $\langle S^z_\tL\rangle$ and $\langle S^z_\tR\rangle $ for asymmetric couplings $\gamma^{(1)}_{\tL \tL}=10\gamma^{(1)}_{\tR \tR}$. The period is $\mathcal{T}=100/(\gamma^{(1)}_{\tL \tL}+\gamma^{(1)}_{\tR \tR})$. Other parameters are identical to Fig.~\ref{fig:symmetric}.}
\end{figure}

So far, we have considered the case where both spins couple symmetrically to the common reservoir. We now turn to the more realistic scenario that the coupling is asymmetric. To begin with, we consider the situation where the left spin couples more strongly than the right spin, $\gamma^{(1)}_{\tL \tL}>\gamma^{(1)}_{\tR \tR}$, and the setup is prepared in the parallel initial state $\ket{\psi_\text{P}}$. Whenever the magnetic field changes direction, the left spin quickly relaxes towards its new ground state, cf. blue curve in Fig.~\ref{fig:asymmetric1}. The right spin (orange curve) behaves differently though. While it shows the tendency to relax to the new ground state upon switching of the magnetic field, it does not fully reach the ground state but rather ends up in a state with $|\langle S^z_\tR\rangle| <S_\tR$. With each switching of the magnetic field, the absolute value of $\langle S^z_\tR\rangle$ decreases such that in the long-time limit a situation is reached where $\langle S^z_\tR\rangle \propto \gamma^{(1)}_\text{RR}/(S_\text{L} \gamma^{(1)}_\text{LL})$ is suppressed. In the limit $\gamma^{(1)}_{\tL \tL} \gg \gamma^{(1)}_{\tR \tR}$ or $S_\text{L}\gg S_\text{R}=1/2$, the expectation value $\langle S^z_\tR\rangle$ vanishes and the system relaxes to an incoherent but evenly distributed mixture of the ground state and $\ket{S^z_\tL=\pm S_\tL$, $S^z_\tR= \mp 1/2}$ for $B(t) \lessgtr 0$. Hence, just like in the case of symmetric coupling, the system reaches a nontrivial state that is different from the ground state and stable until the magnetic field switches again. However, for the asymmetric coupling we find that its properties rather resemble that of an infinite-temperature state (though we would like to emphasize again that the spin state is not a thermal one).

The physics behind this relaxation behavior can be understood most easily in the basis spanned by $S^z_\tL$ and $S^z_\tR$. The left spin relaxes much faster than the right spin due to the stronger coupling strength. During this relaxation, the exchange field mixes degenerate states with $S^z_\tL\pm1$ and $S^z_\tR\mp1$. Moreover, the coherent mixing hinders the thermalization of $S^z_\tR$. After some switches of the magnetic field, no spin direction is preferred and the right spin ends up in an incoherent mixture of being in the ground and in the excited state.

For the case of an antiparallel initial state $\ket{\psi_\text{AP}}$, the system behaves in a similar manner and, before each switching of the magnetic field, reaches a state in which $S^z_\tR$ is suppressed (dotted black curve in Fig.~\ref{fig:asymmetric1}).

\begin{figure}
    \includegraphics[width=\columnwidth]{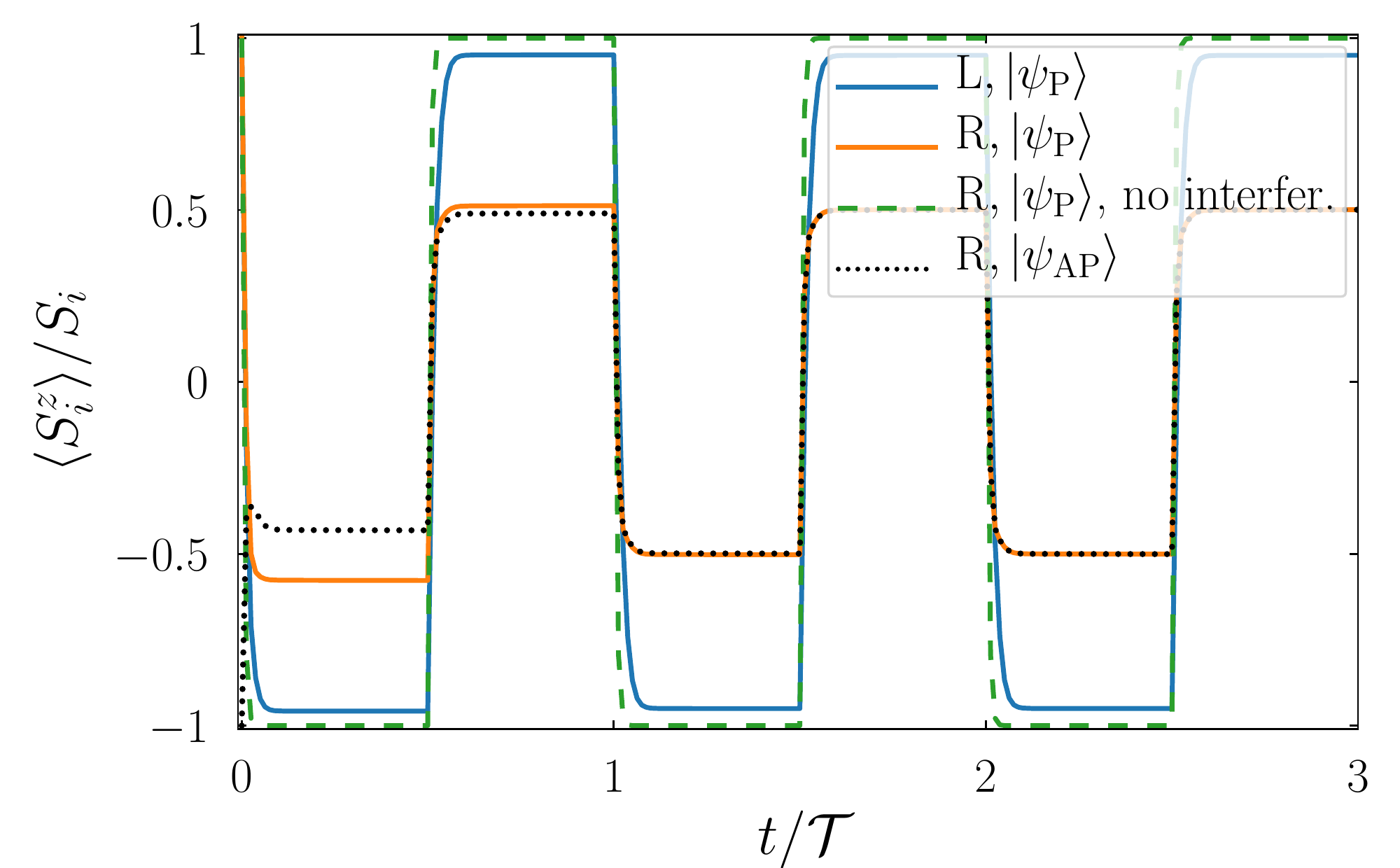}
	\caption{\label{fig:asymmetric2}Time evolution of $\langle S^z_\tL\rangle $ and $\langle S^z_\tR\rangle$ for asymmetric couplings $10\gamma^{(1)}_{\tL \tL}=\gamma^{(1)}_{\tR \tR}$. The period is $\mathcal{T}=100/(\gamma^{(1)}_{\tL \tL}+\gamma^{(1)}_{\tR \tR})$. Other parameters are identical to Fig.~\ref{fig:symmetric}.}
\end{figure}

Finally, we consider the case where the right spin is coupled more strongly to the common reservoir, $\gamma^{(1)}_{\tR \tR}>\gamma^{(1)}_{\tL \tL}$ and the system is prepared in the parallel initial state $\ket{\psi_\text{P}}$. As can be seen in Fig.~\ref{fig:asymmetric2}, in this scenario the spin dynamics is close to what one would expect naively. Upon switching the direction of the magnetic field, both spins flip from their respective excited to ground state. However, we remark that even here due to the presence of coherent superpositions between different spin states and the action of the exchange field the second spin does not reach its ground state completely but has a certain admixture of excited states in the stationary state resulting in a nonvanishing spin expectation value $|\langle S^z_\tR\rangle|<S_\tR$. A similar behavior also occurs when the system is initialized in the antiparallel configuration $\psi_\text{AP}$ (dotted black curve in Fig.~\ref{fig:asymmetric2}). For $S_\text{R}=1/2$, the system relaxes either to its ground state or, with the same probability, to a coherent mixture of $\ket{S^z_\tL=\pm S_\tL$, $S^z_\tR= \mp 1/2}$ and $\ket{S^z_\tL=\pm S_\tL \mp 1$, $S^z_\tR= \pm 1/2}$ for $B(t) \lessgtr 0$~\footnote{In the limit $\gamma^{(1)}_\text{RR} / \gamma^{(1)}_\text{LL} \to \infty$, the coherent mixture decays to the pure state ${\ket{S^z_\tL=\pm S_\tL \mp 1 \text{, } S^z_\tR= \pm 1/2}}$. For $S_\text{L} \to \infty$, the mixture decays to ${\ket{S^z_\tL=\pm S_\tL \text{, }S^z_\tR= \mp 1/2}}$. In the former case, ${\langle S^z_\text{R} \rangle=0}$ and ${\langle S^z_\text{L} \rangle=\pm S_\tL}$. In the latter case, ${\langle S^z_\text{R} \rangle=\pm 1/2}$ but ${\langle S^z_\text{L} \rangle=\pm S_\tL \mp 1/2}$.}.

\subsection{\label{ssec:quench}Quench dynamics}
\begin{figure}
	\includegraphics[width=\columnwidth]{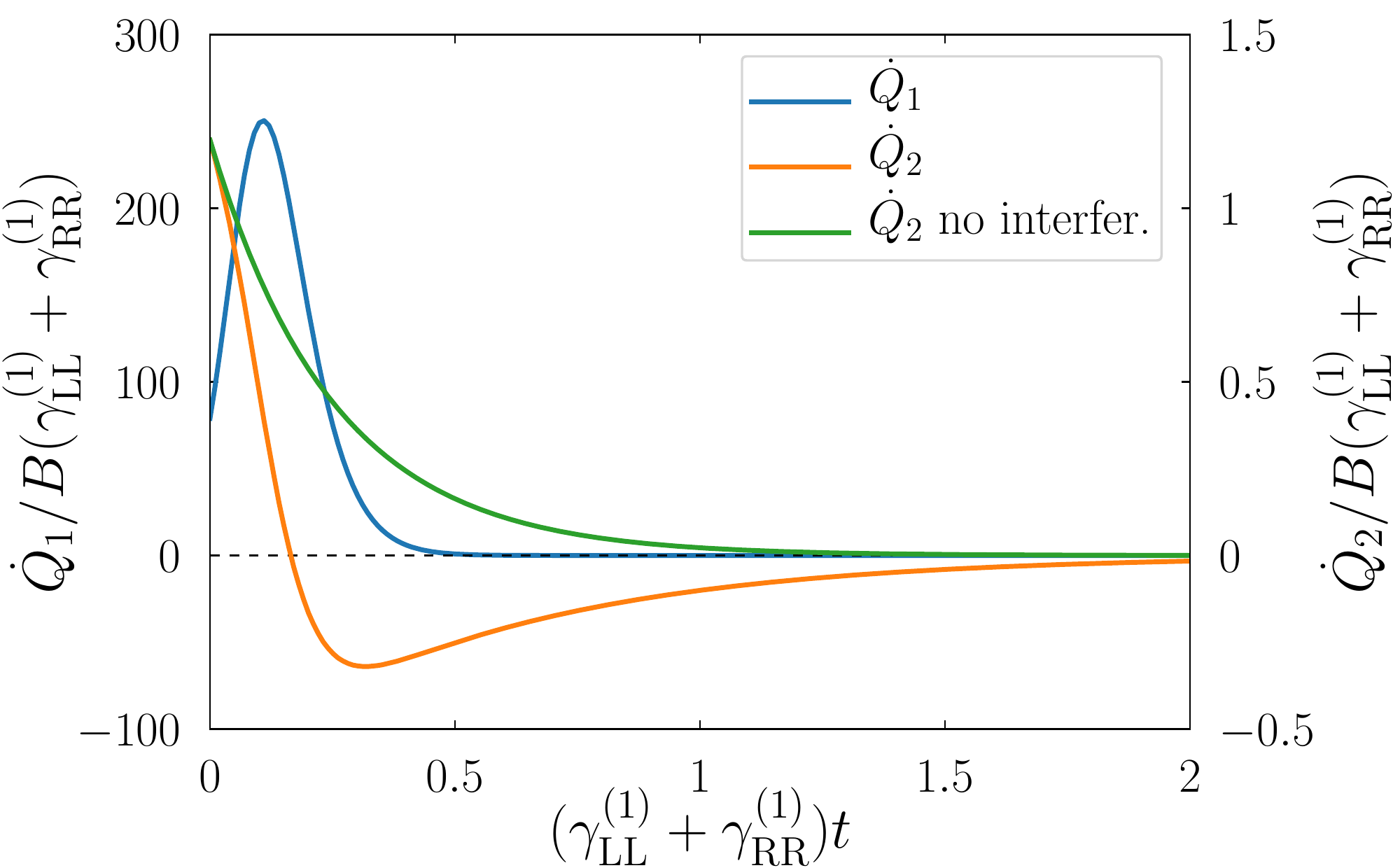}
	\caption{\label{fig:Qdot}Heat currents into the reservoirs as a function of time. The total spins of the spins are $S_\tL=25$ and $S_\tR=1/2$. Both reservoirs are at equal temperature $T$. A magnetic field of strength $B=0.5\, k_\text{B} T$ is switched on at $t=0$. Other parameters are $\gamma^{(1)}_{\tL \tL}=\gamma^{(1)}_{\tR \tR}=\gamma^{(2)}_{\tR \tR}$ and $\omega_\text{c}=10^2 k_\text{B} T$.}
\end{figure}

So far, we have considered the situation where both spins have been coupled to a common reservoir only and where the external magnetic field has been switched periodically. Now, we turn to the case where both spins couple symmetrically to the common reservoir, $\gamma^{(1)}_{\tL \tL}=\gamma^{(1)}_{\tR \tR}$, while the right spin is in addition coupled to a second bosonic reservoir with coupling strength $\gamma^{(2)}_{\tR \tR}$ and consider the dynamics after a quench of the magnetic field.

To this end, we prepare the system in the parallel initial state $\ket{\psi_\text{P}}$. At time $t=0$, the external magnetic field is switched on to point along the negative $z$ direction, $B(t>0)=-B$ such that the spin system is in its highest excited state.
At short times after the quench, the spins relax by creating bosons in reservoir 1 with a rate that scales as $(S_\tL+S_\tR)^2 \gamma^{(1)}_{\tL \tL}$. At the same time, the relaxation via the creation of bosons in the second reservoir scales as $S_\tR^2 \gamma^{(2)}_{\tR \tR}$ and, therefore, is much slower since $S_\tL \gg S_\tR$. As a result, the majority of the heat released during the relaxation process is injected into reservoir 1 via the positive heat current $\dot Q_1$, see Fig.~\ref{fig:Qdot}. Details on the derivation of the heat current can be found in the Appendix~\ref{ap:heat}.

Due to the symmetric coupling of both spins to the first reservoir, the corresponding transitions cannot change the total spin of the spins and states with a total spin $S_\text{tot}<S_\tL+S_\tR$ are not accessible. They become accessible, however, via coupling to the second reservoir which breaks the conservation of the total spin. In consequence, spin flips of $\vec S_\tR$ occur which extract energy  from the second reservoir and lead to a negative heat current $\dot Q_2(t)$, cf. Fig.~\ref{fig:Qdot}. When integrating the heat current $\dot Q_2(t)$ over time during the whole relaxation processes, we find that in total energy is extracted from reservoir~2, i.e., it is cooled down during the relaxation of the spin system. Similar to Sec.~\ref{ssec:periodic}, the importance of the interference effects due to the coupling to the common bath becomes clear when they are neglected manually in the calculation. In this case, $\dot Q_2(t)$ does not change sign as a function of time and reservoir 2 simply is heated, see the green curve in Fig.~\ref{fig:Qdot}.

The amount of heat that can be extracted from the second reservoir can be tuned by the ratio between the coupling to the two reservoirs and can reach about 40\% of $2BS_\tR$. Similarly, the efficiency of the cooling processes which we define as the ratio between the heat extracted from reservoir 2 and the total energy released into the reservoirs during the relaxation process can reach values of about a few percent. While this low efficiency limits the practical use of the proposed refrigeration mechanism, we emphasize that it is nevertheless an interesting effect which demonstrates that a coherent system-bath coupling can give rise to unexpected physical effects.

\section{\label{sec:con}Conclusions}
We have investigated the dynamics of two magnetic spins coupled to bosonic baths.
We found that interference effects due to the coupling of both spins to the same bath gives rise to a nontrivial spin dynamics when the system is driven by the periodic switching of an external magnetic field. In particular, the system does not simply relax to the ground state but rather reaches a long-time behavior that sensitively depends on the initial state as well as the coupling asymmetry between the two spins.
Furthermore, we have demonstrated that the interference effects allow for the realization of a magnetic refrigerator where heat is extracted from one bosonic reservoir during the relaxation processes of the spins while the excess energy is dumped into another reservoir.

In perspective, our work demonstrates that a rich and nontrivial dynamics can be expected whenever several quantum systems are coupled coherently to common thermal baths.

\begin{acknowledgments}
We acknowledge financial support by the German Research Foundation (DFG) within the Collaborative Research Centre (SFB) 1242 ``Non-Equilibrium Dynamics of Condensed Matter in the Time Domain'' (Project No. 278162697) and from the Ministry of Innovation NRW via the ``Programm zur Förderung der Rückkehr des hochqualifizierten Forschungsnachwuchses aus dem Ausland''.
\end{acknowledgments}


\appendix
\section{\label{ap:sol}Solution of the master equation}
In this Appendix, we discuss the generalized master equation~(\ref{eq:master}) in detail. The first term on the right-hand side vanishes since coherences arise between states with different $S_\text{tot}$ but equal $S^z_\text{tot}$. The remaining dissipative term is evaluated under the Born-Markov approximation. Thus, the generalized master equation takes the form $\dot \rho_{\chi_1,\chi_2}(t)=\sum_{\chi'_1,\chi'_2}W_{\chi_1,\chi_2;\chi'_1,\chi'_2}\, \rho_{\chi'_1,\chi'_2}(t)$. The transition rates $W_{\chi_1,\chi_2;\chi'_1,\chi'_2}$ depend on the direction of the applied magnetic field, i.e., either $B(t)=+B$ or $B(t)=-B$. They are evaluated up to first order in the system-reservoir coupling strengths $\gamma^{(j)}_{ii'}$. The contributions of the reservoir $j=1,2$ are independent of each other and, therefore, can be summed up. We concentrate on the first reservoir in the following. Processes emitting a boson into the reservoir (\text{in}) and processes absorbing a boson from the reservoir (\text{out}) lead to the transition rates
\begin{equation}\label{eq:W+}
W^{\text{in}}_{\chi_1,\chi_2;\chi'_1,\chi'_2}= n^-(B)e^{-\frac{B}{\omega_\text{c}}}   \langle \chi_2' | S^{\sigma}_\text{int}|  \chi_2 \rangle   \langle \chi_1 | S^{-\sigma}_\text{int}  |  \chi_1' \rangle \, ,
\end{equation}
\begin{equation}\label{eq:W-}
W^{\text{out}}_{\chi_1,\chi_2;\chi'_1,\chi'_2}= n^+(B)e^{-\frac{B}{\omega_\text{c}}}   \langle \chi_2' | S^{-\sigma}_\text{int}|  \chi_2 \rangle   \langle \chi_1 | S^{\sigma}_\text{int} |  \chi_1' \rangle \, ,
\end{equation}
with $S^{-}_\text{int}=\sqrt{\gamma_\text{LL}^{(1)}} S_\text{L}^{-}+\sqrt{\gamma_\text{RR}^{(1)}} S_\text{R}^{-}$ and $S^{+}_\text{int}=\sqrt{\gamma_\text{LL}^{(1)}} S_\text{L}^{+}+\sqrt{\gamma_\text{RR}^{(1)}} S_\text{R}^{+}$. Moreover, virtual transitions must be taken into account where a boson is exchanged only for an intermediate time span between spins and reservoir. After the transition, the reservoir is back in its initial state, but coherences between states of different $S_\text{tot}$ have been built up or destroyed. The corresponding transition rates take the form
\begin{equation}\label{eq:W0}
\begin{aligned}
&W^{\text{vir}}_{\chi_1,\chi_2;\chi'_1,\chi'_2}=\\
&-\frac{1}{2}\left[n^-(B)e^{-\frac{B}{\omega_\text{c}}}   + \frac{i }{\pi}B^{-}_\text{ex}\right]\bra{\chi'_2} S^{\sigma}_\text{int} S^{-\sigma}_\text{int} \ket{\chi_2} \delta_{\chi_1\chi'_1}\\
&-\frac{1}{2}\left[n^+(B)e^{-\frac{B}{\omega_\text{c}}}  - \frac{i}{\pi}B^{+}_\text{ex}\right]\bra{\chi'_2} S^{-\sigma}_\text{int} S^{\sigma}_\text{int} \ket{\chi_2} \delta_{\chi_1\chi'_1}\\
&-\frac{1}{2}\left[n^-(B)e^{-\frac{B}{\omega_\text{c}}}   - \frac{i}{\pi}B^{-}_\text{ex}\right]\delta_{\chi_2' \chi_2}\bra{\chi'_1} S^{\sigma}_\text{int} S^{-\sigma}_\text{int} \ket{\chi_1}\\
&-\frac{1}{2}\left[n^+(B)e^{-\frac{B}{\omega_\text{c}}}  + \frac{i}{\pi}B^{+}_\text{ex}\right]\delta_{\chi_2'\chi_2}\bra{\chi_1} S^{-\sigma}_\text{int} S^{\sigma}_\text{int} \ket{\chi_1'}\, ,
\end{aligned}
\end{equation}
with exchange-field parameters
\begin{equation}\label{eq:Bpar}
\begin{aligned}
B^{-}_\text{ex}&=-\frac{\omega_\text{c}}{B} + \text{Ei}\left(\frac{B}{\omega_\text{c}}\right) e^{-\frac{B}{\omega_\text{c}}} +B^{+}_\text{ex}\, ,\\
B^{+}_\text{ex}&=\text{Re} \int_{0}^\infty \frac{d\omega}{B} \frac{\omega \, n^{+}(\omega)e^{-\omega/\omega_\text{c}}}{B-\omega + i 0^+}\, .
\end{aligned}
\end{equation}
Here, $\text{Ei}(x)$ is the exponential integral.
The transition rates for the second reservoir are given by Eqs.~(\ref{eq:W+}), (\ref{eq:W-}), and (\ref{eq:W0}) with the replacements $\gamma_\text{LL}^{(1)} \to 0$ and $\gamma_\text{RR}^{(1)}\to \gamma_\text{RR}^{(2)}$.

Finally, we arrange the density matrix elements in a vector~$\bm{\rho}$ and the transition rates in a matrix~$\bm{W}^{(j)}$ for each reservoir. The solution of the master equation takes the form $\bm{\rho}(t)=\exp[{(\bm{W}^{(1)}+\bm{W}^{(2)})(t-t')}]\bm{\rho}(t')$ during a time interval of constant magnetic field. The initial density matrix $\bm{\rho}(t')$ is the final density matrix of the previous time interval with magnetic field of opposite direction.

\section{\label{ap:heat}Heat current}
To obtain the heat current, we must evaluate the heat current kernel $\bm{W}^{(j)}_Q$. It is obtained form~$\bm{W}^{(j)}$ by multiplying each rate by the amount of energy emitted into or absorbed from the reservoir $j$. The procedure is similar to the introduction of counting factors in the field of full counting statistics~\cite{stegmann_detection_2015,stegmann_short-time_2016, luo_stochastic_2019}. Thus, the rates $W^{\text{in}/\text{out}}_{\chi_1,\chi_2;\chi'_1,\chi'_2}$ are multiplied by $\pm B$ and $W^{\text{vir}}_{\chi_1,\chi_2;\chi'_1,\chi'_2}$ by $0$. We obtain the heat current as $Q_j(t)= \text{Tr}[\bm{W}^{(j)}_Q \bm{\rho}(t)]$ where the trace sums up all diagonal density-matrix elements.


\end{document}